\documentstyle[12pt,fullpage]{article}

\newcommand{\Sec}[1]{Section~\ref{#1}}
\newcommand{\eq}[1]{Eq.\,(\ref{eq#1})}
\newcommand{\cv}{{\cal V}}
\newcommand{\cw}{{\cal W}}

\title{Quantum Zeno-like effect and spectra of particles in cascade
transition}
\author{Alexander D.\ Panov\\
{\small Skobeltsyn Institute of Nuclear Physics,}\\
{\small Moscow State University, Moscow, Russia}
\thanks{E-mail address: {\tt a.panov@relcom.ru}}
}
\date{}

\begin{document}
%%%%%%%%%%%%%%%%%%%%%%%%%%%%%%%%%%%%%%%%%%%%%%%%%%%%%%%%%%%%%%%%%%%
%\bibliographystyle{prsty}
%%%%%%%%%%%%%%%%%%%%%%%%%%%%%%%%%%%%%%%%%%%%%%%%%%%%%%%%%%%%%%%%%%%

\maketitle

\centerline{PACS numbers: 03.65.Bz; 32.70.Jz}
\centerline{Keywords: quantum measurement theory, Zeno effect,
spontaneous decay}

\begin{abstract}
Shr\"odinger equation for two-step spontaneous cascade transition in
a three-level quantum system is solved by means of Markovian
approximation for non-Markovian integro-differential evolution
equations for amplitudes of states. It is shown that both decay
constant and radiation shift of initial level are affected by
instability of intermediate level of the cascade. These phenomena are
interpreted as the different manifestations of quantum Zeno-like
effect.  The spectra of particles emitted during the cascade
transition are calculated in the general case and, in particular, for
an unusual situation when the initial state is lower than the
intermediate one. It is shown that the spectra of particles do not
have a peak-like shape in the latter case.
\end{abstract}

\section{Introduction}
\label{INTRODUCTION}

The term ``quantum Zeno paradox'' had been introduced in
\cite{ZEN_SUDARSHAN77A,ZEN_SUDARSHAN77B}.  It was argued there that
an unstable particle which was continuously observed in order to see
whether it decays would never be found to decay  (for review see
\cite{ZEN_HOME97}). In the present paper we restrict our
consideration to the special case of continuous waiting-mode (or
negative result) observations of spontaneous decay.  An example of
such measurements is a registration, by a permanently presented
detector, of particles emitted during quantum state decay. Until the
detector is ``discharged'', we continuously obtain information that
the system is in the initial excited state. Another interesting
example of waiting-mode observation of spontaneous decay one can find
in \cite{ZEN_ELATTARI99}. It was shown \cite{PANOV96F} that in the
general case decay may be ``frozen'' by continuous waiting-mode
observation only at the limit of infinitely fast reaction of
measuring device on event of decay (infinitely short decoherence
time).  However, in case of realistic decoherence time the decay may
be perturbed in various directions:  it may be either slowed down or
fastened. The sign of the effect depends on details of transition
matrix elements behavior and on transition energy.  This is a quantum
Zeno effect, not a paradox.  Thus, in our definition, the quantum
Zeno effect is any influence of continuous measurement to probability
of decay. At the limit of infinitely fast measuring device quantum
Zeno effect means ``freezing'' of decay.  Qualitatively, the
complicated behavior of quantum Zeno effect may be related to a
complicated behavior of initial part of decay curve.  We will discuss
this relation in more details elsewhere.

It is a difficult problem to determine how a permanently presented
detector affects on the probability of decay in realistic situations
\cite{PANOV96F}. But the kinematics of this process are very similar
to those of a group of phenomena which was called quantum Zeno-like
effects \cite{PANOV99B}. All these phenomena (including above
mentioned observation) are described by the same general equation for
perturbation of decay probability \cite{PANOV99B}. Also, these
phenomena has a common main feature:  The final state of decay could
not be considered as stable, but further transition to other
orthogonal states occur. Just these transitions perturb the decay
constant. The consideration of some of Zeno-like effects turns out to
be much simpler than the genuine Zeno effect.  So, it is reasonable
to begin with more simple problems.

Systems that show Zeno-like effects differ from each other by the
reason of transition from final state of decay.  The system with
forced resonance transition from the final state of decay was studied
in \cite{ZEN_PASCAZIO97,ZEN_PASCAZIO99A,ZEN_PASCAZIO99B}.  The
analogous system was considered as a particular case of a general
Zeno-like system in \cite{PANOV99B}.  M.~B.~Mensky
\cite{ZEN_MENSKY99A} was the first who proposed to consider a
spontaneously decaying system with a spontaneously decaying final
state as the system demonstrating quantum Zeno or Zeno-like effect.
A simple example of such a system is the system with two-step
spontaneous cascade transition. Such systems are the subject of the
present paper.

Let $X$ be a three-level system (for example, atom or atomic nucleus)
with states $|x_0\rangle$, $|x_1\rangle$, $|x_2\rangle$ and
eigenenergies $\omega^x_0$, $\omega^x_1$, $\omega^x_2$ respectively
(it is assumed $\hbar = 1$ hereafter).  Suppose that system $X$ was
prepared in state $|x_0\rangle$ at the initial moment of time $t=0$.
State $|x_0\rangle$ is unstable and decays spontaneously to state
$|x_1\rangle$ due to interaction with another system (``field''). The
latter system has a continuous spectrum of states.  Let state
$|x_1\rangle$ be also unstable. System $X$ further decays from state
$|x_1\rangle$ to final stable state $|x_2\rangle$. We suppose for
simplicity that direct transition from state $|x_0\rangle$ to
$|x_2\rangle$ is forbidden. Such a system exhibits cascade
spontaneous transition from state $|x_0\rangle$ to state
$|x_2\rangle$ through intermediate state $|x_1\rangle$. This
phenomenon was studied in some details many years ago
\cite{DEC_WEISSKOPF30A,DEC_ROSENFELD31,DEC_CASIMIR33} and was
discussed in classical monographs \cite{HEITLER36,BERESTETSKY82}. A
new property of cascade transition that was pointed out in
\cite{ZEN_MENSKY99A} was that the instability of level $|x_1\rangle$
should affect the life-time of level $|x_0\rangle$.

It was noted in \cite{ZEN_MENSKY99A} that during cascade transition
the second fast transition $|x_1\rangle\to |x_2\rangle$ after decay
of initial state $|x_0\rangle$ to $|x_1\rangle$ was similar to
waiting-mode observation of first decay $|x_0\rangle\to |x_1\rangle$.
The main difference of the second transition from a genuine
continuous measurement is that it is not possible to switch off the
interaction causing the second transition, but it is possible to stop
a measurement.  Hence, the perturbation of decay rate of transition
$|x_0\rangle\to |x_1\rangle$ by instability of state $|x_1\rangle$
may be attributed to quantum Zeno-like effect.  We use the term
quantum Zeno effect as synonym for quantum Zeno-like effect throughout
the present paper.

The formula for decay rate of state $|x_0\rangle$ perturbed by
instability of state $|x_1\rangle$ was derived in \cite{PANOV99B}.
With notations analogous to those in our paper, this formula reads as
\begin{equation}
   \widetilde\Gamma_0 = 2\pi \int_0^\infty d\omega \cv(\omega)
   \frac{1}{\pi}
   \frac{\lambda_1}{\lambda_1^2 + (\omega -\omega_{01}+\mu_1)^2}\,\,.
   \label{eq1}
\end{equation}
Here $\omega_{01}=\omega^x_0-\omega^x_1$; $\cv(\omega)$ is the sum of
all square modula of transition matrix elements related to the same
energy $\omega$ of emitted particle; $\lambda_1$ is the real part of
decay constant of level $|x_1\rangle$; $\mu_1$ is the contribution to
radiation shift of level $|x_1\rangle$ from discrete level
$|x_2\rangle$ \cite{DEC_SEKE94B}. We shall hereafter mention similar
energy shifts as radiation shifts simply. The complex decay constant
$\gamma_1$ of level $|x_1\rangle$ is $\gamma_1 = \lambda_1 + i\mu_1$
if the system was prepared in state $|x_1\rangle$.  At the limit of
$\lambda_1\to 0$ \eq{1} transforms into conventional Fermi's Golden
rule:
\begin{equation}
   \Gamma_0 = 2\pi\cv(\omega_{01} - \mu_1),
   \label{eq2}
\end{equation}
but wherein transition energy is corrected by radiation shift of
level $|x_1\rangle$. However, deviation of perturbed probability
$\widetilde\Gamma_0$ from unperturbed value $\Gamma_0$ exists, if
$\lambda_1 \ne 0$. Just this phenomenon is considered as quantum Zeno
effect in \cite{ZEN_MENSKY99A}. It is easily seen from \eq{1} that
Zeno effect is strong if $\lambda_1$ is comparable with $\omega_{01}$
by its value. If formally $\lambda_1\to\infty$ we obtain
$\widetilde\Gamma_0=0$.  This is pure ``quantum Zeno paradox''.

It is possible to make an interesting conclusion from \eq{1}.
Suppose $\omega_{01}-\mu_1 < 0$. Then Golden rule \eq{2} predicts
zero probability of decay of level $|x_0\rangle$ since $\cv(\omega)
\equiv 0$ for $\omega < 0$. There are no field quanta with negative
energy. However, \eq{1} shows that perturbed value of probability of
decay $\widetilde\Gamma_0$ is grater than zero in this situation
generally.  This phenomenon is a special case of quantum Zeno effect.
Hence, the transition from lower ($|x_0\rangle$) to upper
($|x_1\rangle$) level is possible, and positive energy quanta should
be emitted in such a process.  So, an important question arises:
What are the spectra of quanta emitted during transitions
$|x_0\rangle\to |x_1\rangle$ and $|x_1\rangle\to |x_2\rangle$ in this
unusual situation? Also, what are these spectra in the general case
when transition energy $\omega_{01}$ is comparable with decay
constant $\lambda_1$ of level $|x_1\rangle$?  Obviously, these
spectra cannot be Lorentzian-shape peaks.

\eq{1} was derived on the base of the second order perturbation theory
in \cite{PANOV99B}. The spectra of emitted particles can not be
calculated by this method. So, the present paper includes two aims.
Firstly, we derive no-decay amplitude of initial state $|x_0\rangle$
and probability $\widetilde\Gamma_0$ by nonperturbative method which
is based on direct transition from non-Markovian evolution equations
to Markovian approximation.  Secondly, we obtain all spectra of
interest from our nonperturbative solution of Shr\"odinger equation:
mutual energy distribution of particles emitted in the first and
second transitions, spectra of particles emitted in the first and
second transitions separately, and distribution of the sum of energy
of first and second emitted particles.  Some of these spectra was
determined early \cite{BERESTETSKY82}, but those results are not
related to the conditions $\omega_{01} \sim \lambda_1$ or
$\omega_{01} < 0$.

This paper is organized as follows.  In \Sec{MARKOV} we discuss the
Markovian approximation for a spontaneous exponential decay in
two-level system.  In \Sec{CASCADE} we use the formalism developed in
\Sec{MARKOV} for description of cascade transition in a three-level
system and obtain our main results:  perturbed values of decay
constant and radiation shift of level $|x_0\rangle$ and spectra of
particles.  Finally, some features of these results are discussed in
\Sec{DISCUSSION} and conclusions are drawn.

\section{Markovian approximation in two-level problem}
\label{MARKOV}

We consider a model of spontaneous transition of a general type. Let
$X$ be a two-level system $(|x_0\rangle,|x_1\rangle)$. System $X$
interacts with another system $F$ (field). System $F$ has a ground
state $|f_0\rangle$ and continuous spectrum of exited states
$|y_\eta\rangle$, where $\eta$ is the index of state in the
continuous spectrum. Since we discuss in \Sec{CASCADE} a field with
quanta of two different kinds, we use the notation $|y\rangle$ for
field quanta instead of $|f\rangle$. The continuous spectrum states
are normalized by condition
$
   \langle y_\eta|y_{\eta'} \rangle = \delta(\eta-\eta').
$
Let the initial state of the combined system $X\otimes F$ at time
$t=0$ be
\begin{displaymath}
   |\Psi(0)\rangle = |x_0\rangle \otimes |f_0\rangle \equiv |x_0
   f_0\rangle.
   %\label{eq3}
\end{displaymath}
The Hamiltonian of system is
\begin{displaymath}
   H = H_0 + V
   %\label{eq4}
\end{displaymath}
where $H_0$ is a ``free'' Hamiltonian
\begin{displaymath}
   H_0 = \omega_0^x|x_0\rangle\langle x_0| +
         \omega_1^x|x_1\rangle\langle x_1| +
         \int \omega_\eta^y b^+_\eta b_\eta d\eta
   %\label{eq5}
\end{displaymath}
and $V$ is an interaction between $X$ and $F$:
\begin{equation}
   V = \int \left[v(\eta)b^+_\eta|x_1\rangle\langle x_0| +
             v^*(\eta)b_\eta |x_0\rangle\langle x_1|\right] d\eta.
   \label{eq6}
\end{equation}
In \eq{6} $b^+_\eta$ is the creation operator for state
$|y_\eta\rangle$. It does not matter what is the commutation relation
for operators $b_\eta$: either
$[b_\eta,b^{+}_{\eta'}]_{-} = \delta(\eta-\eta')$
or
$[b_\eta,b^{+}_{\eta'}]_{+} = \delta(\eta-\eta')$.
It is easy to see that $v(\eta)$ is a matrix element of transition:
$v(\eta) = \langle x_1 y_\eta|V|x_0 f_0\rangle$.

To solve the Shr\"odinger equation
\begin{equation}
   |\dot\Psi(t)\rangle = -i(H_0 + V)|\Psi(t)\rangle
   \label{eq7}
\end{equation}
we use the ansatz
\begin{equation}
   |\Psi(t)\rangle = |x_0 f_0\rangle a_0(t)e^{-i\omega^x_0 t} +
   \int d\eta\,|x_1 y_\eta\rangle a_{1\eta}(t)
   e^{-i(\omega^x_1 + \omega^y_\eta)t}.
   \label{eq8}
\end{equation}
Substituting \eq{8} for $|\Psi(t)\rangle$ in \eq{7} we obtain the
system of equations:
\begin{eqnarray}
   \dot a_0(t) & = & -i\int d\eta\,v^*(\eta)
   e^{-i(\omega^y_\eta - \omega_{01})t} a_{1\eta}(t)
   \label{eq9}\\
   \dot a_{1\eta}(t)& = & -iv(\eta) e^{i(\omega^y_\eta - \omega_{01})t}
   a_0(t).
   \label{eq10}
\end{eqnarray}
\eq{10} can be solved for coefficients $a_{1\eta}$. Substituting the
solution for $a_{1\eta}$ in \eq{9} we get the equation for coefficient
$a_0(t)$:
\begin{equation}
   \dot a_0(t) = -\int_0^t a_0(t_1) q_0(t-t_1) dt_1,
   \label{eq11}
\end{equation}
where
\begin{equation}
   q_0(\tau) = \int |v(\eta)|^2 e^{-i(\omega^y_\eta -
   \omega_{01})\tau} d\eta.
   \label{eq12}
\end{equation}
The amplitude $a_0(t)$ is a solution of non-Markovian equation
(\ref{eq11}): the derivative of $\dot a_0$ at the moment of time $t$
is expressed through all values of $a_0$ for all moments of time from
$0$ to $t$.

A.~Sudbery \cite{ZEN_SUDBERY84,ZEN_SUDBERY86} proposed qualitative
arguments that the function $q_0(\tau)$ in \eq{11} was a very narrow
peak around value $\tau = 0$ for usual decay systems. Besides, it is
possible to understand why it should be the case if we consider the
behavior of function $v(\eta)$.

The index $\eta$ of state $|y_\eta\rangle$ can be represented as the
eigenenergy of the state $\omega^y$ and the degeneration index
$\alpha^y$: $\eta = \{\omega^y,\alpha^y\}$. Then \eq{12} can be
rewritten as
\begin{equation}
   q_0(\tau) = e^{i\omega_{01}\tau}
   \int_0^\infty \cv(\omega^y)e^{-i\omega^y\tau} d\omega^y,
   \label{eq13}
\end{equation}
where
\begin{equation}
   \cv(\omega^y) = \int |v(\omega^y,\alpha^y)|^2d\alpha^y.
   \label{eq14}
\end{equation}
The integral in \eq{14} means a sum for discrete indexes.  \eq{13}
shows that the function $q_0(\tau)$ is a Fourier transform of
function $\cv(\omega^y)$ up to factor $\exp(i\omega_{01}\tau)$ which
is equal to one by module. The function $\cv(\omega^y)$ is very wide
for usual decay systems. For example, in the case of electromagnetic
2P-1S transition of hydrogen atom, the value $\Lambda$ of natural
cut-off of function $\cv(\omega^y)$ is $\Lambda = \frac{3}{2}\alpha
m_e \approx 5.6\cdot10^3$\,eV, where $\alpha$ is the fine structure
constant and $m_e$ is the electron mass
\cite{DEC_MOSES73,DEC_SEKE94A,ZEN_PASCAZIO98}. The value of $\Lambda$
is much greater than the energy of 2P-1S transition. The Fourier
transform of wide real non-negative function $\cv(\omega^y)$ is a
narrow peak near $\tau = 0$. Consequently, the function $q_0(\tau)$
is a narrow peak near $\tau = 0$ too. The width of this peak is about
$\tau_{Zen} = 1/\Lambda$.  Suppose $a_0(t)$ to vary slowly during
time intervals of the order of $1/\omega_{01}$ for times $t \gg
1/\omega_{01}$.  Then, for the same times, $a_0(t)$ is approximately
constant during the time interval of order $\tau_{Zen}$ and the
function $a_0(t-t_1)$ may be moved out from the integral in \eq{11}
at time $t$.  Making also the variable change $\tau = t-t_1$ we
rewrite \eq{11} as
\begin{equation}
   \dot a_0(t) = -a_0(t)\int_0^t q_0(\tau)d\tau.
   \label{eq15}
\end{equation}
Therefore, we obtain an approximate Markovian equation (\ref{eq15})
for $a_0(t)$ instead of non-Markovian equation (\ref{eq11}). Recall
that \eq{15} is valid only for $t >> 1/\omega_{01}$. It is not
difficult to calculate the integral in r.h.s.\ of \eq{15} using
\eq{13}:
\begin{equation}
   \int_0^t q_0(\tau)d\tau =
   \int_0^\infty \cv(\omega^y)
   \left[
   \frac{\sin(\omega^y-\omega_{01})t}{\omega^y-\omega_{01}} -
   i\frac{1-\cos(\omega^y-\omega_{01})t}{\omega^y-\omega_{01}}
   \right]
   d\omega^y.
   \label{eq16}
\end{equation}
Suppose $\cv(\omega^y)$ is sufficiently smooth. Then it is seen that
the integral in the r.h.s.\ of \eq{16} does not depend on time for $t
>> 1/\omega_{01}$. Hence, we can change the upper limit of integral
from $t$ to infinity and find that
\begin{displaymath}
   \int_0^\infty q_0(\tau)d\tau = \gamma_0 = \lambda_0 + i\mu_0,
   %\label{eq17}
\end{displaymath}
where
\begin{eqnarray}
   \lambda_0  & = &  {\rm Re}\,\gamma_0 = \pi \cv(\omega_{01})
   \nonumber
   %\label{eq18}
   \\
   \mu_0  & = & {\rm Im}\,\gamma_0 = -P\int_0^\infty
   \frac{\cv(\omega^y)}{\omega^y - \omega_{01}} d\omega^y.
   \label{eq19}
\end{eqnarray}
Here $P$ denotes the principal value of an integral.

\eq{15} reads now as $\dot a_0(t) = -\gamma_0 a_0(t)$ and has the
solution $a_0(t) = \exp(-\gamma_0 t)$. This is the usual exponential
decay law. The real part of $\gamma_0$ determines the probability of
decay per unit of time $\Gamma_0 = 2{\rm Re}\,\gamma_0 =
2\pi\cv(\omega_{01})$; the imaginary part of $\gamma_0$ is the
radiation shift of level $|x_0\rangle$.

\section{Decay constants, radiation shifts, and spectra in
two-step cascade transition}
\label{CASCADE}

We consider three-level system $X$ with cascade transition
$|x_0\rangle\to |x_1\rangle\to |x_2\rangle$ in this section. The
transitions result from interaction of system $X$ with another system
$F$ (field).  Suppose that two different types of quanta are emitted
during the first and during the second transition. The quanta
$|y_\eta\rangle$ are created during transition $|x_0\rangle\to
|x_1\rangle$ and the quanta $|z_\zeta\rangle$ are created during
transition $|x_1\rangle\to |x_2\rangle$. The $y$ and $z$ quanta have
creation operators $b^+_\eta$ and $c^+_\zeta$, respectively. We admit
that for all $\eta$ and $\zeta$ the operators $b_\eta$ and $c_\zeta$
satisfy the relation
\begin{equation}
  [b_\eta,c^+_\zeta]_-=0,\quad
  [b_\eta,c_\zeta]_-=0.
  \label{eq20}
\end{equation}
\eq{20} represents the meaning of difference between particles $y$
and $z$. Operators $b_\eta$ may be either Bozonic or Fermionic type,
the same is true for operators $c_\zeta$. The statistics type of $y$
and $z$ particles may be different from each other. For example, we
may consider cascade nuclear transition when an atomic electron is
emitted in the fist transition (the inner nuclear conversion
phenomenon) and electromagnetic quantum is emitted in the second one.
Our model is correct for this case. We also can consider a cascade
electromagnetic transition, but the energy of the first transition is
much less than the energy of the second one. In this case \eq{20} is
not strictly true for all quanta of such cascade transition, but our
model may be accounted as a good approximation in this case as well.
The generalization to the case when we cannot distinguish between
quanta emitted in the first transition and in the second transition
is not straightforward and is not discussed here.

The Hamiltonian of the system $X \otimes F$ is
\begin{displaymath}
   H = H_0 + V + W,
   %\label{eq21}
\end{displaymath}
where
\begin{eqnarray}
   H_0 &=&
   \sum_{\xi = 0}^2 \omega^x_\xi |x_\xi\rangle\langle x_\xi| +
   \int d\eta\,\omega^y_\eta b^+_\eta b_\eta +
   \int d\zeta\,\omega^z_\zeta c^+_\zeta c_\zeta,
   \label{eq22}\\
   V &=&
   \int d\eta\,
   \left[
   v(\eta)b^+_\eta |x_1 \rangle\langle x_0| +
   v^*(\eta) b_\eta |x_0 \rangle\langle x_1|
   \right],
   \label{eq23}\\
   W &=&
   \int d\zeta\,
   \left[
   w(\zeta)c^+_\zeta |x_2 \rangle\langle x_1| +
   w^*(\zeta) c_\zeta |x_1 \rangle\langle x_2|
   \right].
   \label{eq24}
\end{eqnarray}
The notations in Eqs.\,(\ref{eq22}--\ref{eq24}) are similar to
those of \Sec{MARKOV} and obvious.  For initial state
$|\Psi(0)\rangle = |x_0 f_0\rangle$ we solve the Shr\"odinger
equation
\begin{equation}
   |\dot \Psi(t)\rangle = -i(H_0+V+W)|\Psi(t)\rangle
   \label{eq25}
\end{equation}
using the ansatz
\begin{eqnarray}
  |\Psi(t)\rangle &=& |x_0 f_0\rangle a_0(t) e^{-i\omega^x_0 t} +
  \int d\eta\,|x_1 y_\eta\rangle a_{1\eta}(t)
  e^{-i(\omega^x_1 + \omega^y_\eta)t}
  \nonumber \\
  & &
  + \int d\eta \int d\zeta\, |x_2 y_\eta z_\zeta\rangle
  a_{2\eta\zeta}(t)
  e^{-i(\omega^x_2 + \omega^y_\eta + \omega^z_\zeta)t}.
  \label{eq26}
\end{eqnarray}
Substituting \eq{26} for $|\Psi(t)\rangle$ in \eq{25} we obtain the
system of equations
\begin{eqnarray}
   \dot a_0(t) & = & -i\int d\eta\,v^*(\eta)
   e^{-i(\omega^y_\eta - \omega_{01})t} a_{1\eta}(t)
   \label{eq27}\\
   \dot a_{1\eta}(t) & = & -iv(\eta)
   e^{i(\omega^y_\eta - \omega_{01})t} a_0(t) -
   i\int d\zeta\,w^*(\zeta)
   e^{-i(\omega^z_\zeta - \omega_{12})t} a_{2\eta\zeta}(t)
   \label{eq28}\\
   \dot a_{2\eta\zeta}(t) & = & -iw(\zeta)
   e^{i(\omega^z_\zeta - \omega_{12})t} a_{1\eta}(t).
   \label{eq29}
\end{eqnarray}
Here $\omega_{ij} = \omega^x_i - \omega^x_j$.  From \eq{29} we have
\begin{equation}
   a_{2\eta\zeta}(t) = -iw(\zeta)\int_0^t dt_1 \,
   e^{i(\omega^z_\zeta - \omega_{12})t_1}
   a_{1\eta}(t_1).
   \label{eq30}
\end{equation}
Substituting \eq{30} for $a_{2\eta\zeta}(t)$ in \eq{28},
we obtain a non-Markovian equation for coefficients $a_{1\eta}$:
\begin{equation}
   \dot a_{1\eta}(t) = -iv(\eta) e^{i(\omega^y_\eta-\omega_{01})t}
   a_0(t) -
   \int_0^t dt_1\,a_{1\eta}(t_1)q_1(t-t_1),
   \label{eq31}
\end{equation}
where
\begin{eqnarray}
   q_1(\tau) & = & e^{i\omega_{12}\tau}
   \int \cw(\omega^z) e^{-i\omega^z\tau} d\omega^z,
   \label{eq32}\\
   \cw(\omega^z) & = & \int |w(\omega^z, \alpha^z)|^2 d\alpha^z
   \label{eq33}
\end{eqnarray}
and we supposed $\zeta = \{\omega^z, \alpha^z\}$.  The equations
(\ref{eq32},\ref{eq33}) are quite similar to equations
(\ref{eq13},\ref{eq14}), \Sec{MARKOV}. The only difference is that
Eqs.\,(\ref{eq13},\ref{eq14}) are related to transition
$|x_0\rangle\to |x_1\rangle$ but Eqs.\,(\ref{eq32},\ref{eq33}) are
related to transition $|x_1\rangle\to |x_2\rangle$. The integral in
r.h.s.\ of \eq{31} is similar to the integral in r.h.s.\ of \eq{11}.
Therefore, arguing as in \Sec{MARKOV}, we see that \eq{31} may be
changed by approximate Markovian equation
\begin{equation}
   \dot a_{1\eta}(t) =
   -iv(\eta)e^{i(\omega^y_\eta-\omega_{01})t} a_0(t) -
   \gamma_1 a_{1\eta}(t),
   \label{eq34}
\end{equation}
where
\begin{eqnarray*}
   &&\gamma_1 = \lambda_1 +i\mu_1 = \int_0^\infty q_1(\tau)d\tau,\\
   &&\lambda_1 = \pi\cw(\omega_{12});\quad
   \mu_1 = -P\int_0^\infty
   \frac{\cw(\omega^z)}{\omega^z-\omega_{12}} d\omega^z.
\end{eqnarray*}
The solution of \eq{34} is
\begin{equation}
   a_{1\eta}(t) = -iv(\eta)
   \int_0^t e^{-\gamma_1(t-t_1)} e^{i(\omega^y_\eta-\omega_{01})t_1}
   a_0(t_1).
   \label{eq35}
\end{equation}
Substituting \eq{35} for $a_{1\eta}(t)$ in \eq{27} we find the
equation for amplitude $a_0(t)$:
\begin{equation}
   \dot a_0(t) = -\int_0^t a_0(t_1) \tilde q_0(t-t_1) dt_1,
   \label{eq36}
\end{equation}
where
\begin{equation}
   \tilde q_0(\tau) =
   e^{-\gamma_1\tau} e^{i\omega_{01}\tau}
   \int \cv(\omega^y) e^{-i\omega^y\tau} d\omega^y.
   \label{eq37}
\end{equation}
The tilde indication of function $\tilde q_0(\tau)$ means that this
function is related to transition $|x_0\rangle\to |x_1\rangle$
perturbed by instability of state $|x_1\rangle$. Further the meaning
of tilde will be the same in all cases. Function $\tilde q_0(\tau)$
differs from nondisturbed function $q_0(\tau)$ \eq{13} by additional
factor $\exp(-\gamma_1 \tau)$. The module of this factor is a
decreasing function since ${\rm Re}\,\gamma_1 = \lambda_1 > 0$.
Consequently, the function $\tilde q_0(\tau)$ is a narrow peak near
the value $\tau = 0$ as well as the nondisturbed function $q_0(\tau)$
(see \Sec{MARKOV}). Hence, we can change the non-Markovian equation
(\ref{eq36}) to Markovian one
\begin{equation}
   \dot a_0(t) = -\tilde \gamma_0 a_0(t),
   \label{eq38}
\end{equation}
where
\begin{equation}
   \tilde \gamma_0 = \tilde \lambda_0 + i \tilde \mu_0 =
   \int_0^\infty \tilde q_0(\tau) d\tau.
   \label{eq39}
\end{equation}
It is not difficult to obtain from \eq{39} and \eq{37}:
\begin{eqnarray}
   \tilde \lambda_0 & = & \pi
   \int_0^\infty \cv(\omega^y) \frac{1}{\pi}
   \frac{\lambda_1}
   {\lambda_1^2 + (\omega^y - \omega_{01} + \mu_1)^2} d\omega^y,
   \label{eq40}\\
   \tilde \mu_0 & = &
   \int_0^\infty \cv(\omega^y)
   \frac{\omega^y - \omega_{01} + \mu_1}
   {\lambda_1^2 + (\omega^y - \omega_{01} + \mu_1)^2} d\omega^y.
   \label{eq41}
\end{eqnarray}
Solving \eq{38}, we get
\begin{equation}
   a_0(t) = e^{-\tilde \gamma_0 t}.
   \label{eq42}
\end{equation}
It follows from \eq{42} that $\tilde \gamma_0$ is the complex decay
constant of state $|x_0\rangle$. Decay constant is perturbed by
instability of state $|x_1\rangle$. Thus, the probability of decay
per unit of time is $\widetilde\Gamma_0 = 2{\rm Re}\,\tilde\gamma_0 =
2\tilde\lambda_0$.  This value coincides with the result obtained
early in \cite{PANOV99B} by perturbation method (comp.\ \eq{1} and
\eq{40}).

Now let us find the spectra of particles $y$ and $z$ created during
the first and second transitions of system $X$. These spectra are
defined by values $|a_{1\eta}(t)|^2$ and $|a_{2\eta\zeta}(t)|^2$ as
$t\to\infty$. Substituting \eq{42} for $a_0(t)$ in \eq{35}, we obtain
\begin{equation}
   a_{1\eta}(t) = -iv(\eta)
   \frac
   {
      e^{[i(\omega^y_\eta-\omega_{01})-\tilde\gamma_0]t} -
      e^{-\gamma_1 t}
   }
   {
      i(\omega^y_\eta - \omega_{01}) + \gamma_1 - \tilde\gamma_0
   }.
   \label{eq43}
\end{equation}
It is readily seen that
\begin{displaymath}
   \lim_{t\to\infty} |a_{1\eta}(t)|^2 = 0.
   %\label{eq44}
\end{displaymath}
This means that coefficients $a_{1\eta}(t)$ do not contribute to
spectra of particles. This could be expected because these
coefficients relate to intermediate state of the system.

Substituting \eq{43} for $a_{1\eta}(t)$ in \eq{30}, we get the
expression for $a_{2\eta\zeta}(t)$:
\begin{eqnarray}
   a_{2\eta\zeta}(t) & = &
   -\frac{v(\eta)w(\zeta)}
   {i(\omega^y_\eta-\omega_{01}) + \gamma_1 - \tilde\gamma_0}
   \nonumber\\
   &&
   \times
   \left\{
   \frac
   {
      e^{[i(\omega^y_\eta + \omega^z_\zeta -\omega_{02})
      -\tilde\gamma_0]t} - 1
   }
   {
     i(\omega^y_\eta + \omega^z_\zeta -\omega_{02})
     -\tilde\gamma_0
   }
   -
   \frac
   {
      e^{[i(\omega^z_\zeta - \omega_{12}) - \gamma_1]t} - 1
   }
   {
      i(\omega^z_\zeta - \omega_{12}) - \gamma_1
   }
   \right\}.
   \label{eq45}
\end{eqnarray}
It is easy to obtain from \eq{45} that the limit of
$a_{2\eta\zeta}(t)$ as $t\to\infty$ is
\begin{equation}
   a_{2\eta\zeta}(\infty) =
   \frac
   {
      v(\eta)w(\zeta)
   }
   {
      [i(\omega^y_\eta+\omega^z_\zeta-\omega_{02}) - \tilde\gamma_0]
      [i(\omega^z_\zeta-\omega_{12})-\gamma_1]
   }.
   \label{eq46}
\end{equation}
Now we can calculate mutual distribution of energy of particles
$y$ and $z$:
\begin{equation}
   p(\omega^y,\omega^z) = \int d\alpha^y \int d\alpha^z\,
   |a_{2;\omega^y\alpha^y;\omega^z\alpha^z}(\infty)|^2.
   \label{eq47}
\end{equation}
>From \eq{46} and \eq{47} we get
\begin{equation}
   p(\omega^y,\omega^z) =
   \frac
   {
      \cv(\omega^y)\cw(\omega^z)
   }
   {
      [\tilde\lambda^2_0 + (\omega^y+\omega^z-\bar\omega_{02})^2]
      [\lambda_1^2 + (\omega^z-\bar\omega_{12})^2]
   }.
   \label{eq48}
\end{equation}
where $\bar\omega_{02}$ and $\bar\omega_{12}$ are the corrected
values of transition energies
\begin{displaymath}
   \bar\omega_{02} = (\omega^x_0 + \tilde\mu_0) - \omega^x_2;
   \quad
   \bar\omega_{12} = (\omega^x_1 + \mu_1) - \omega^x_2.
   %\label{eq49}
\end{displaymath}
Let us note that the energy $\omega^x_0$ is corrected by perturbed
value of radiation shift $\tilde\mu_0$ defined by \eq{41} instead of
unperturbed radiation shift \eq{19}.

The spectrum of particles $y$ created in the first transition is
defined by
\begin{equation}
   p_y(\omega^y) =
   \int_0^{+\infty} p(\omega^y,\omega^z)d\omega^z =
   \int_{-\infty}^{+\infty} p(\omega^y,\omega^z)d\omega^z.
   \label{eq50}
\end{equation}
We change the lower limit of integral in \eq{50} from 0 to
$-\infty$ since $\cw(\omega^z) = 0$ for all $\omega^z < 0$. It is
possible to calculate the integral \eq{50} analytically only if the
function $\cw(\omega^z)$ is known. In the general case we have to
introduce some approximation. Suppose that $|\omega_{01}| \ll
\omega_{12}$, $\omega^y \ll \omega_{12}$, and $\cw(\omega^z)$ is
a sufficiently smooth function:
\begin{equation}
   \cw(\omega^y + \bar\omega_{02}) \approx
   \cw(\bar\omega_{12}) \approx
   \cw(\omega_{12}).
   \label{eq51}
\end{equation}
Then we can rewrite \eq{50} as
\begin{equation}
   p_y(\omega^y) = \cv(\omega^y) \cw(\omega_{12})
   \int_{-\infty}^{+\infty}
   \frac{d\omega^z}
   {
      [\tilde\lambda^2_0 + (\omega^y+\omega^z-\bar\omega_{02})^2]
      [\lambda_1^2 + (\omega^z-\bar\omega_{12})^2]
   }.
   \label{eq52}
\end{equation}
It is not hard to calculate the integral in \eq{52} by residue
theory. We obtain:
\begin{equation}
   p_y(\omega^y) =
   \frac{\pi\cv(\omega^y)}{\tilde\lambda_0}
   \left[
   \frac{1}{\pi}
   \frac
   {\tilde\lambda_0 + \lambda_1}
   {(\tilde\lambda_0 + \lambda_1)^2 + (\omega^y-\bar\omega_{01})^2}
   \right],
   \label{eq53}
\end{equation}
where $\bar\omega_{01} = (\omega^x_0+\tilde\mu_0) -
(\omega^x_1+\mu_1)$ is the corrected energy of transition
$|x_0\rangle\to|x_1\rangle$.

The spectrum of particles $z$ is
\begin{eqnarray}
   p_z(\omega^z) &=& \int_0^{+\infty} p(\omega^y,\omega^z) d\omega^y
   \nonumber\\
   &=&
   \frac{\cw(\omega^z)}{\lambda_1^2+(\omega^z-\bar\omega_{12})^2}
   \int_{-\infty}^{+\infty}
   \frac
   {\cv(\omega^y) d\omega^y}
   {\tilde\lambda_0^2 + (\omega^y+\omega^z-\bar\omega_{02})^2}.
   \label{eq54}
\end{eqnarray}
Suppose $\cv(\omega^y)$ to vary slowly during intervals of order
$\tilde\lambda_0$ for all $\omega^y$. Then we can move
$\cv(\omega^y)$ out of integral in \eq{54} for
$\omega^y=\bar\omega_{02}-\omega^z$. Taking into account also \eq{51}
we have
\begin{equation}
   p_z(\omega^z) =
   \frac{\pi\cv(\bar\omega_{02}-\omega^z)}{\tilde\lambda_0}
   \left[
   \frac{1}{\pi}
   \frac
   {\lambda_1}
   {\lambda_1^2+(\omega^z-\bar\omega_{12})^2}
   \right].
   \label{eq55}
\end{equation}

Let us find the distribution of the sum of energies $\omega^y +
\omega^z = \Omega$. It can easily be checked that
\begin{equation}
   p_{y+z}(\Omega) = \int_0^\Omega
   p(\omega^y,\Omega-\omega^y) d\omega^y.
   \label{eq56}
\end{equation}
It follows from \eq{56} and \eq{48} with assumption \eq{51} that
\begin{equation}
   p_{y+z}(\Omega) =
   \frac
   {S(\Omega)}
   {\tilde\lambda_0^2 + (\Omega-\bar\omega_{02})^2},
   \label{eq57}
\end{equation}
where
\begin{displaymath}
   S(\Omega) = \int_0^\Omega \cv(\omega^y)
   \left[
   \frac{1}{\pi}
   \frac
   {\lambda_1}
   {\lambda_1^2 + (\Omega-\omega^y-\bar\omega_{12})^2}
   \right]
   d\omega^y.
   %\label{eq58}
\end{displaymath}
If $\lambda_1 \gg \tilde\lambda_0$, function $S(\Omega)$ varies
slowly in comparison with the pole-like denominator of \eq{57}. Hence,
the spectrum of the sum of particle $y$ and $z$ energies is
approximately a narrow Lorentzian-shape peak of width
$\tilde\lambda_0$ (as could be expected).

\section{Discussion and conclusions}
\label{DISCUSSION}

The main results of the present paper are following:
\begin{itemize}
\item
\eq{40} describes perturbed value of the real part of decay constant
of level $|x_0\rangle$ (the initial level of cascade transition).
The real part is also the half of decay probability per unit of time
of level $|x_0\rangle$.
\item
\eq{41} describes perturbed value of imaginary part of decay constant
of level $|x_0\rangle$. The imaginary part is the perturbed value of
radiation shift of level $|x_0\rangle$.
\item
\eq{48} describes mutual energy spectrum of particles of the first
and of the second transition of a cascade.
\item
\eq{53} describes energy spectrum of the first transition of a
cascade.
\item
\eq{55} describes energy spectrum of the second transition of a
cascade.
\item
\eq{57} describes distribution of the sum of particle energies
created during the first and the second transitions of a cascade.
\end{itemize}

We discussed \eq{1} for perturbed value of decay probability
$\widetilde\Gamma_0$ in the Introduction. Since $\widetilde\Gamma_0 =
2\tilde\lambda_0$, so this discussion is related to \eq{40} as well.

\eq{41} shows that instability of level $|x_1\rangle$ affects the
discrete level contribution to radiation shift of level $|x_0\rangle$
as well as the probability of decay. Therefore, the well-known
formula for radiation shift (\ref{eq19}) should be replaced by
\eq{41} if $\lambda_1$ is comparable with $|\omega_{01}|$.  It is
easy to see that \eq{41} transforms into usual \eq{19} as
$\lambda_1\to 0,\mu_1\to 0$.  If formally $\lambda_1\to\infty$, from
\eq{41} we obtain $\tilde\mu_0\to 0$.  This result is similar to
$\tilde\lambda_0\to 0$ as $\lambda_1\to \infty$, therefore it may be
called ``an energy-shift quantum Zeno paradox''.  Similarly, the
perturbation of radiation shift of level $|x_0\rangle$ by instability
of level $|x_1\rangle$ for realistic values $\lambda_1$ may be called
``an energy-shift quantum Zeno effect". Note that it could be
expected that ``energy-shift quantum Zeno effect'' would be presented
in waiting-mode observation of decay in the general case, not only in
cascade transitions. Thus, the same mechanism that perturbs the
probability of decay also perturbs the radiation shift of level.

Let us now discuss the expression for particle spectra emitted during
the first transition (\eq{53}) and during the second transition
(\eq{55}).  It is suitable to discuss tree different situations:

$1^{\rm o}$. If $\lambda_1 \ll \bar\omega_{01}$ and $\bar\omega_{01}
> 0$ then it could be considered that function $\cv(\omega^y)$ to
vary very slowly in comparison with the pole-like denominators in
\eq{53} and \eq{55}.  Hence, we obtain that the spectra defined by
Eqs.~(\ref{eq53},\ref{eq55}) are usual Lorentzian-shape peaks. The
width of the spectra of first transition is $\tilde\lambda_0 +
\lambda_1$, but not $\tilde\lambda_0$. These conclusions are quite
similar to well-known results \cite{BERESTETSKY82}, but $\lambda_0$
in \cite{BERESTETSKY82} is now changed by perturbed value
$\tilde\lambda_0$.

$2^{\rm o}$. If $\lambda_1\sim \bar\omega_{01}$ and $\bar\omega_{01}
> 0$, it can not be considered that function $\cv(\omega^y)$ to vary
slowly in comparison with the denominators in \eq{53} and \eq{55}.
Therefore, both the spectra of particles $y$ and $z$ become
strongly deformed Lorentzian peaks.

$3^{\rm o}$. Finally, suppose $\bar\omega_{01} < 0$. Then the maxima
of Lorentzian factors of \eq{53} and \eq{55} are positioned in the
branch of $\omega$ values where $\cv(\omega) = 0$. The spectra
shapes are defined by the shape of function $\cv(\omega)$ multiplied
by the tale of Lorentzian peaks now.  Therefore, both spectra
$p_y(\omega^y)$ and $p_z(\omega^z)$ are continuous rather than
peak-like. The energy of quanta $y$ emitted during the first step of
cascade transition is positive, of course, in spite of
$\bar\omega_{01} < 0$.

Thus, the separate spectra of particles $y$ and $z$ may be very wide
or even continuous. But it follows from \eq{57} that the energies of
particles $y$ and $z$ remain strongly correlated such that the width
of the distribution of sum $\Omega = \omega^y + \omega^z$ is equal to
$\tilde\lambda_0$ in all cases. This is a manifestation of
fundamental uncertainty principle for energy and time.

\vskip 0.5cm
\centerline{\bf ACKNOWLEDGMENTS}

The author acknowledges the fruitful discussions with M.~B.~Mensky
and V.~A.~Namiot and is grateful to V.~A.~Aref'ev for the help in
preparation of the paper. The work was supported in part by the
Russian Foundation of Basic Research, grant 98-01-00161.

%%%%%%%%%%%%%%%%%%%%%%%%%%%%%%%%%%%%%%%%%%%%%%%%%%%%%%%%%%%%%%%%%
%\bibliography{../quant}
%%%%%%%%%%%%%%%%%%%%%%%%%%%%%%%%%%%%%%%%%%%%%%%%%%%%%%%%%%%%%%%%%

\end{document}